\documentclass[12pt,preprint]{aastex}
\usepackage{emulateapj5}           

\slugcomment{Draft from \today, submitted to ApJ}

\shorttitle{SMA 440$\mu$m/690GHz observations of Orion-KL}
\shortauthors{Beuther et al.}

\begin{document}


\title{Submillimeter Array 440$\mu$m/690GHz line and continuum observations
of Orion-KL}


\author{H.~Beuther$^1$, Q.~Zhang$^1$, M.J.~Reid$^1$, T.R. Hunter$^1$,
M. Gurwell$^1$, D.~Wilner$^1$, J.-H.~Zhao$^1$, H.~Shinnaga$^1$,
E.~Keto$^1$, P.T.P.~Ho$^1$, J.M.~Moran$^1$, S.-Y.~Liu$^2$}

\altaffiltext{1}{Harvard-Smithsonian Center for Astrophysics, 60 Garden Street, Cambridge, MA 02138, USA}
\altaffiltext{2}{Academia Sinica Institute of Astronomy and Astrophysics,  National Taiwan University, No.1, Roosevelt Rd, Sec. 4, Taipei 106, Taiwan, R.O.C.}

\email{hbeuther@cfa.harvard.edu}








\begin{abstract} 
Submillimeter Array observations of Orion-KL at $\sim 1''$ resolution
in the 440\,$\mu$m/690\,GHz band reveal new insights about the
continuum and line emission of the region. The 440\,$\mu$m continuum
flux density measurement from source {\it I} allows us to
differentiate among the various proposed physical models: Source {\it
I} can be well modeled by a ``normal'' protostellar SED consisting of
a proton-electron free-free emission component at low frequencies and
a strong dust component in the submillimeter bands. Furthermore, we
find that the protostellar object SMA1 is clearly distinct from the
hot core. The non-detection of SMA1 at cm and infrared wavelengths
suggests that it may be one of the youngest sources in the entire
Orion-KL region. The molecular line maps show emission mainly from the
sources {\it I}, SMA1 and the hot core peak position. An analysis of
the CH$_3$CN$(37_K-36_K)$ $K$-ladder ($K=0...3$) indicates a warm gas
component of the order $600\pm 200$\,K. In addition, we detect a large
fraction ($\sim 58\%$) of unidentified lines and discuss the
difficulties of line identifications at these frequencies.
\end{abstract}


\section{Introduction}
\label{intro}

The 440$\mu$m/690GHz band is almost entirely unexplored at high
spatial resolution. In February 2005, the Submillimeter Array
(SMA\footnote{The Submillimeter Array is a joint project between the
Smithsonian Astrophysical Observatory and the Academia Sinica
Institute of Astronomy and Astrophysics, and is funded by the
Smithsonian Institution and the Academia Sinica.}) performed the first
imaging campaign in that band with 6 antennas in operation, thus line
and continuum imaging at $1''$ resolution was possible at frequencies
around 690\,GHz. One of the early targets of this campaign was the
Orion-KL region.

At a distance of $\sim 450$\,pc, Orion-KL is the closest and most
studied region of massive star formation in our Galaxy. Furthermore,
the Orion-KL hot core is known to be particularly rich in molecular
line emission (e.g.,
\citealt{wright1996,schilke2001,beuther2005a}). The hot core as traced
by (sub)millimeter dust continuum emission is a chain of clumps offset
by $\sim 1''$ from the radio source {\it I}. The hot core is rich in
nitrogen-bearing molecules like CH$_3$CN and NH$_3$ and has
temperatures of the order of a few hundred Kelvin (e.g.,
\citealt{wilner1994,wilson2000,beuther2005a}). In contrast, the
so-called compact ridge approximately $5''$ to the south is
particularly rich in CH$_3$OH emission and has lower temperatures of
the order 100\,K (e.g., \citealt{wright1996,liu2002,beuther2005a}). In
addition, the region harbors a complex cluster of infrared sources
studied from near- to mid-infrared wavelengths (e.g.,
\citealt{dougados1993,gezari1998,greenhill2004,shuping2004}). At least
two outflows are driven from the region, one high-velocity outflow in
the south-east north-west direction (e.g.,
\citealt{allen1993,chernin1996,schultz1999}), and one lower-velocity
outflow in the north-east south-west direction (e.g.,
\citealt{genzel1989,blake1996,chrysostomou1997,stolovy1998,beuther2005a}). The
driving source(s) of the outflows are uncertain. Initial claims that
it might be IRc2 are outdated now, and possible culprits are the radio
source {\it I} and/or the infrared source {\it n}, also known as radio
source {\it L} \citep{menten1995}. 

In 2004, we observed Orion-KL at $1''$ resolution with the SMA in the
865$\mu$m/348GHz band \citep{beuther2004g,beuther2005a}. The molecular
line data show a rich line forest resolving different spatial
structures for the various species. The SiO(8--7) observations
convincingly show that the north-east south-west outflow emission
originates at source {\it I}. Temperature estimates based on a
CH$_3$OH multi-line fit result in values as high as 350\,K for the hot
core. In the 865\,$\mu$m submillimeter continuum data, source {\it I}
was resolved from the hot core and source {\it n} was detected as
well. The spectral energy distribution of source {\it I} from 8 to
348\,GHz still allowed ambiguous interpretations of its physical nature
(see also \S \ref{source_i}). Furthermore, a new continuum peak, SMA1,
between sources {\it I} and {\it n} was detected; however, the nature
of SMA1 was uncertain, as it could have been either an independent
protostellar core or part of the extended hot core. In this paper, our
new observations in the 440\,$\mu$m band clarify the nature of SMA1
and source {\it I}.

\section{Observations}
\label{obs}

We observed Orion-KL with the SMA on February 19th, 2005, in the
440$\mu$m/690GHz band. Six antennas were equipped with 690\,GHz
receivers, and the covered baselines ranged between 16 and 68\,m. The
weather conditions were excellent with a zenith opacity, ~-- measured
with the NRAO tipping radiometer located at the Caltech Submillimeter
Observatory~-- $\tau(\rm{230GHz})$ between 0.03 and 0.04 throughout
the night. This corresponds to zenith opacities at 690\,GHz between
0.6 and 0.8 ($\tau(\rm{690GHz})\sim 20\times
(\tau(\rm{230GHz})-0.01)$, \citealt{masson1994}). The phase center was
the nominal position of source {\it I} as given by
\citet{plambeck1995}: R.A.~[J2000] $5^h35^m14.50^s$ and Dec.~[J2000]
$-5^{\circ}22'30''.45$. The FWHM of the primary beam of the SMA at
these frequencies is $18''$, the aperture efficiency is $\sim 0.4$.
For more details on the array and its capabilities see \citet{ho2004},
and recent updates are given at http://sma1.sma.hawaii.edu/specs.html.

The receivers operated in a double-sideband mode with an IF band of
4-6\,GHz so that the upper and lower sidebands were separated by
10\,GHz. The correlator had a bandwidth of 1.968\,GHz and the channel
spacing was 0.8125\,MHz. The covered frequency ranges were 679.78 to
681.75\,GHz and 689.78 to 691.75\,GHz (Fig.~\ref{spectra}). Bandpass
calibration was done with observations of Callisto (angular diameter
$\sim 1.4''$). Since no quasar is strong enough in that band for
calibration purposes, we calibrated phase and amplitude via frequent
observations of Titan (angular diameter $\sim 0.9''$). The distances
of Callisto and Titan from Jupiter and Saturn were $>230''$ and $\sim
190''$, respectively. The contamination of the Callisto and Titan
fluxes by their primaries is negligible due to mainly three reasons:
(a) the primary beam pattern severely attenuates these planets since
they are much more than 10 primary beams away, (b) the large size of
the planets ($40''$ for Jupiter, $20''$ for Saturn) implies that a
large fraction of their fluxes is resolved out, and (c) Due to the
motion of the moons over the period of observations any small
contamination will not be added coherently, further reducing the
contamination issue.

The large angle between Titan and Orion-KL ($\sim 40^{\circ}$)
introduced larger than normal uncertainties in the phase and amplitude
transfer.  The flux calibration was performed with the Titan
observations as well, for which a well-calibrated model has been
created and used extensively at mm wavelengths
\citep{gurwell1995,gurwell2000,gurwell2004}.  In the current
observations, the upper sideband (USB) included the CO(6--5) line,
which is quite broad as determined by using both observations of lower
CO transitions and modeling of the expected line shape, assuming that
CO has a uniform 50 ppm abundance in the Titan atmosphere
\citep{gurwell2004}.  Modeling the flux of the continuum LSB and
continuum+CO(6--5) USB emission, the expected average Titan fluxes at
the given dates were 12.0 and 22.3\,Jy for the LSB and USB,
respectively. Based on flux measurements of the two main continuum
sources in Orion-KL using only the part of the spectra with no
detectable line emission (\S\ref{cont}), we estimate the relative flux
accuracy between the upper and lower sideband to be better than
10\%. The absolute flux calibration was estimated to be accurate
within $\sim 25\%$.

\begin{figure}
\begin{center}
\includegraphics[angle=-90,width=10cm]{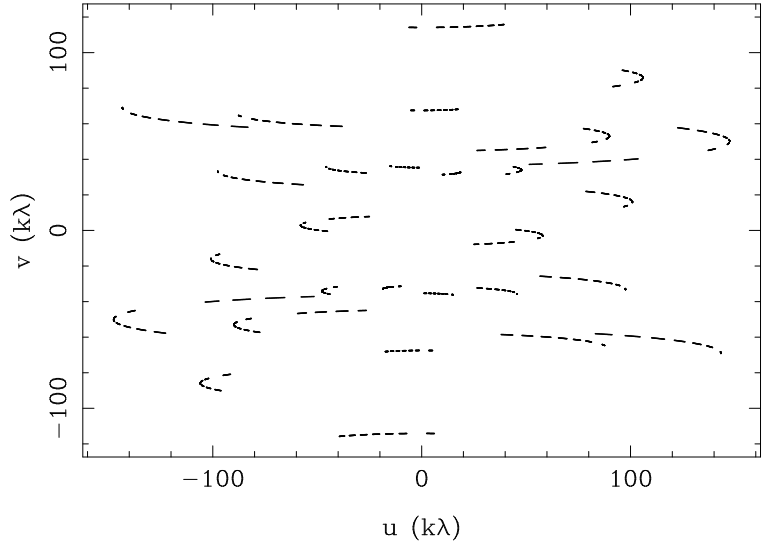}
\end{center}
\caption{The top panel shows the uv-coverage of the observations.
The bottom panel presents the resulting dirty beam, contour levels
cover the range $\pm 1$, in steps of $\pm 0.125$.}
\label{dirty}
\end{figure}

Since Orion-KL is close to the celestial equator, and we have for most
baselines only source data observed in an hour angle range from $-2.6$
to +1.2 hours (the data of only 3 baselines were usable over a longer
range from $-2.6$ to +4.5 hours), imaging of the data proved to be
difficult. Figure \ref{dirty} shows the given uv-coverage and the
resulting dirty beam. In spite of the large-scale emission present in
Orion-KL, we had to employ clean-boxes covering the central region
around sources {\it I}, SMA1 and the main hot core peak to derive
reasonable images because of the high side-lobe levels. Measured
double-sideband system temperatures corrected to the top of the
atmosphere were between 900 and 3500\,K, depending on the elevation of
the source. Our sensitivity was limited by the strong side-lobes of
the strongest emission peaks, and thus varied between the continuum and
the line maps of different molecules and molecular transitions. This
limitation was mainly due to the incomplete sampling of short
uv-spacings and the presence of extended structures. The achieved
$1\sigma$ rms of the 440\,$\mu$m continuum image, produced by
averaging the apparently line-free part of the USB spectrum (see \S
\ref{cont}), was $\sim 700$\,mJy/beam~-- well above the expected
1$\sigma$ of $\sim 41$\,mJy/beam for the given on-source time, the
used bandpass and an average $\tau(230\rm{GHz})=0.037$ (corresponding
to 0.61\,mm precipitable water vapor). This larger rms was mainly due
to the strong side-lobes and inadequate cleaning. The theoretical
$1\sigma$ rms per 1\,km\,s$^{-1}$ channel was $\sim 775$\,mJy/beam,
whereas the measured $1\sigma$ rms in 1\,km\,s$^{-1}$ channel images
was $\sim 2.6$\,Jy/beam, again, because of the strong side-lobes and
inadequate cleaning.  The $1\sigma$ rms for the velocity-integrated
molecular line maps (the velocity ranges for the integrations were
chosen for each line separately depending on the line-widths and
intensities) ranged between 1.1 and 1.6\,Jy/beam. The synthesized
beams were $1.4''\times 0.9''$ using an intermediate weighting between
natural and uniform (``robust'' value 0 in MIRIAD). We calibrated the
data within the IDL superset MIR developed for the Owens Valley Radio
Observatory and adapted for the SMA (\citealt{scoville1993}, see also
the MIR cookbook by Charlie Qi at
http://cfa-www.harvard.edu/$\sim$cqi/micook.html). The imaging was
performed in MIRIAD \citep{sault1995}.

\section{Results and Analysis}

\subsection{440\,$\mu$m continuum emission}
\label{cont}

\begin{figure}
\begin{center}
\includegraphics[angle=-90,width=13cm]{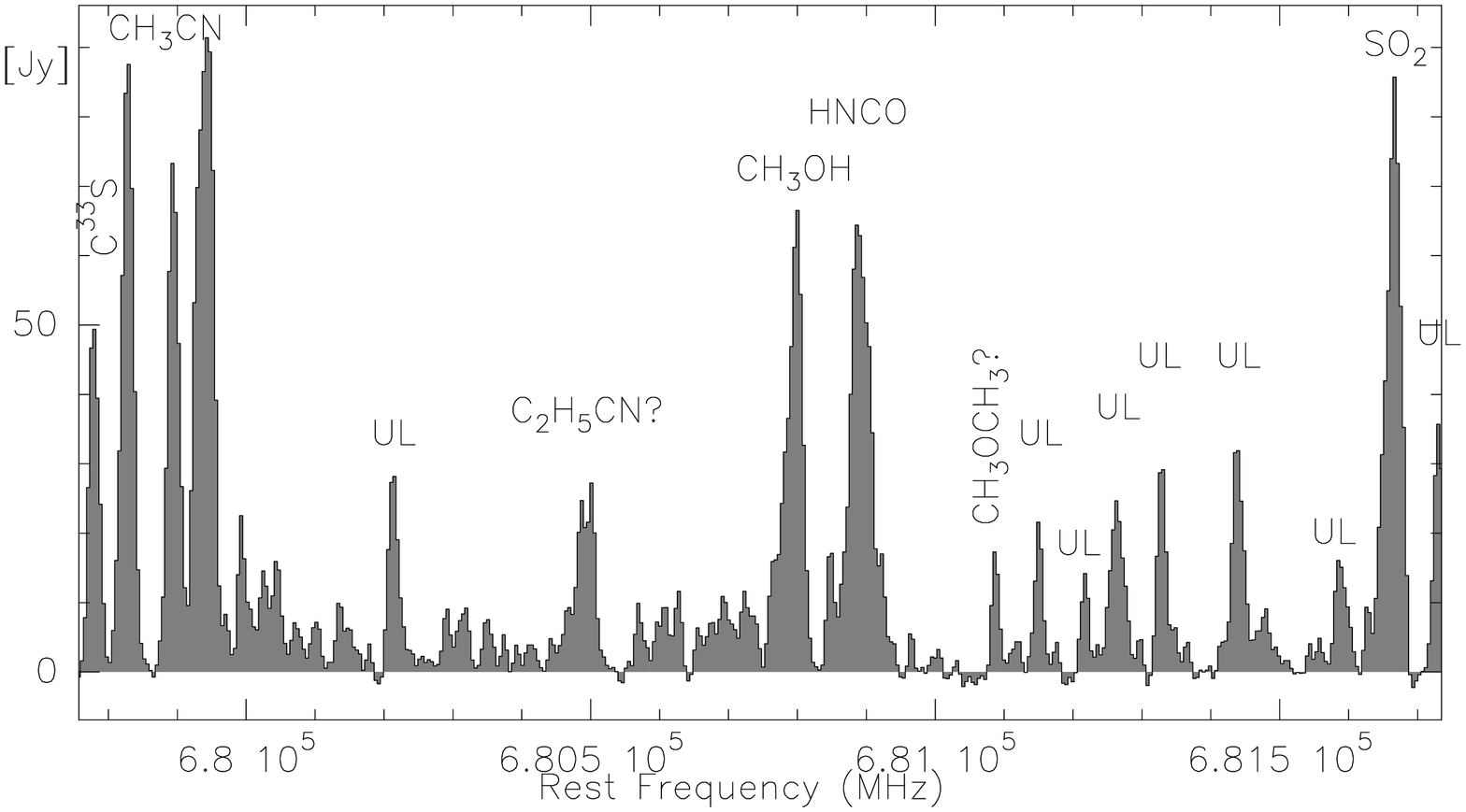}
\includegraphics[angle=-90,width=13cm]{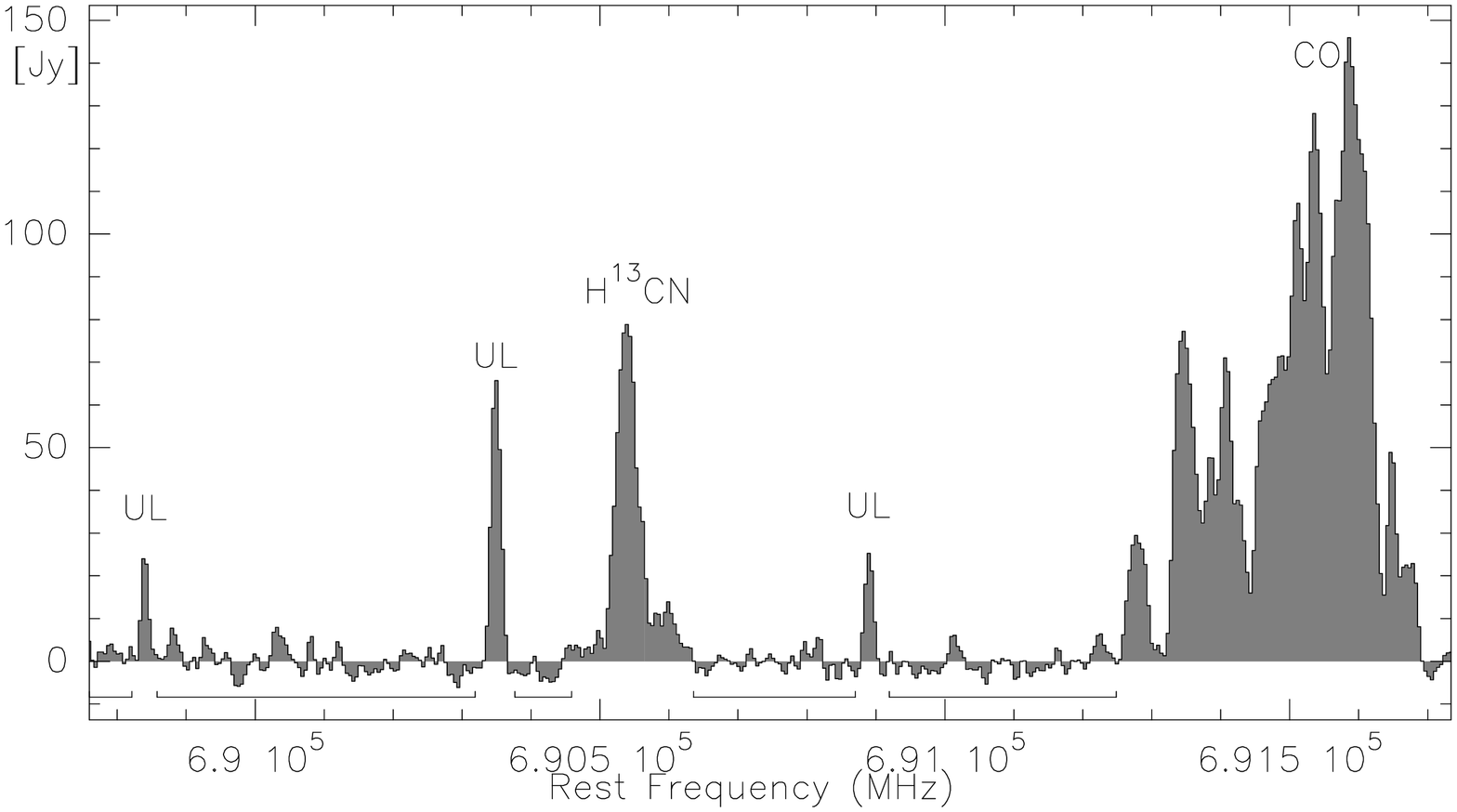}
\end{center}
\caption{Vector-averaged lower and upper sideband spectra in the
uv-domain on a baseline of $\sim 25$\,m. The data are averaged over an
hour angle range from $-2.6$ to +4.5 hours with a total on-source
integration time of 176 minutes. UL marks unidentified lines and ``?''
marks tentatively identified lines. The lines at the bottom of the USB
spectrum mark the apparently line-free part of the spectrum, which we
used to produce the 440\,$\mu$m continuum image (Figs.~\ref{continuum}
\& \ref{model}).}
\label{spectra}
\end{figure}

The spectra presented in Figure \ref{spectra} exhibit many molecular
lines, therefore constructing a continuum image had to be done with great
care. The continuum image presented in Fig.~\ref{continuum} (right
panel) was constructed by averaging the part of the USB spectrum with
no detectable line emission as shown in Fig.~\ref{spectra}. The
corresponding apparently line-free part of the LSB is considerably
smaller and thus the comparable pseudo-continuum image
noisier. Although we therefore omit the LSB data for the scientific
interpretation, we point out that the peak fluxes in the LSB continuum
image vary by less than 10\% compared to the USB image. This shows
the good relative flux calibration between both sidebands.

Weak lines below the detection limit may contribute to the continuum
emission.  To estimate the contamination from such weak lines, we
produced an additional pseudo-continuum dataset of the lower sideband
data including all spectral lines. We used the LSB data for this
purpose because its line-contamination is more severe. Fitting two
point-sources in the uv-domain (a) to the whole LSB continuum data and
(b) to the apparently line-free LSB data (for details see \S
\ref{flux}), the measured fluxes from (a) were
only $\sim 15\%$ higher for source {\it I} and SMA1 than those from
(b).  In our data, possible weak lines unidentified due to noise are
on average less than 10\% of the average peak-flux of the strong lines
(Fig.~\ref{spectra}). Since the strong lines already cover about half
of the given bandpass (Fig.~\ref{spectra}), the contribution from
possible weak lines below the detection limit amounts to only about 1.5\%.
This is negligible compared to the absolute flux uncertainties.

The derived 440\,$\mu$m continuum emission map is shown in the right
panel of Figure \ref{continuum}. To make a proper comparison with the
865\,$\mu$m emission, we also show the original image at that
wavelength \citep{beuther2004g}, as well as an image of the 865\,$\mu$m
observations using only the 30 to 160\,k$\lambda$ data range, which is
the range of baselines sampled by our 440\,$\mu$m data. In spite of
strong side-lobes and the imaging problems described in \S \ref{obs},
the 440\,$\mu$m map shows two distinct point-like sources associated
with source {\it I} and SMA1. Comparing this image with the 30 to
160\,k$\lambda$ image at 865\,$\mu$m, we find at the longer wavelength
similar structures for source {\it I} and SMA1 plus emission from the
hot core. Additional features like source {\it n} are not present in
the 440\,$\mu$m continuum image, likely because of the poor imaging
capabilities and thus strong side-lobes of the observations (see \S
\ref{obs}). The main difference between the 440\,$\mu$m and the
865\,$\mu$m continuum images is the strong presence of the hot core in
the longer wavelength image, whereas we find only a small extension to
the east at 440\,$\mu$m. Imaging of the spectral line data in the
440\,$\mu$m band (\S \ref{line}) shows that, in spite of the mentioned
imaging limitations (\S \ref{obs}), we can spatially image and
separate line emission from source {\it I} and the hot
core. Hence, while it remains puzzling that the typically strong
hot core is only barely detectable in our 440\,$\mu$m image, it might
be a real feature of the data indicating a peculiar low
spectral index $\alpha$ of the hot core emission (for a detailed
discussion see \S\ref{hotcore}).

\subsubsection{Flux measurements}
\label{flux}

The 440\,$\mu$m continuum emission in Fig.~\ref{continuum} gives the
impression that the point sources {\it I} and SMA1 are embedded in a
larger-scale ridge emission. While Orion-KL is known to be part of the
large-scale Orion molecular ridge (e.g., \citealt{lis1998}), the
larger-scale emission in our image is to some degree also an artifact
of the data reduction, since we had to use clean boxes and thus
shifted emission from the entire primary beam to the central
region. Hence, flux measurements toward source {\it I} and SMA1 from
the images will give values that are too high. To avoid these
problems, we fitted the data in the uv-domain assuming:\\ (a) two
point sources at the positions of source {\it I} and SMA1;\\ (b) two
point sources at the positions of source {\it I} and SMA1 + an
elliptical Gaussian;\\ (c) three point sources at the positions of
source {\it I}, SMA1, and the hot core peak position;\\ (d) three
point sources at the positions of source {\it I}, SMA1, and the hot
core peak position + an elliptical Gaussian.\\ Afterwards, we imaged
the models using exactly the same uv-coverage and imaging parameters
employed previously for the real data. Figure \ref{model} shows a
direct comparison between the image of the original data and three
model fits. We find that the models with additional larger-scale
elliptical Gaussian component reproduce the data rather well. Although
the fitted fluxes for source {\it I} and SMA1 vary by less than 15\%
(within the calibration uncertainty) with or without the third point
source at the position of the hot core, the
three-point-sources+Gaussian model image most resembles the original
data image. Therefore, we use this model for the following analysis.

These simplified models allow us to bracket the fluxes from sources
{\it I}, SMA1 and the hot core. The three-point-source-only model
likely overestimates the intrinsic fluxes because it adds some of the
underlying large-scale emission to the point sources. Contrary to
this, the three-point-source+Gaussian model underestimates the
point-source contributions and thus the point-source fluxes. The
measured values for both uv-fitting approaches are:



\begin{tabbing}
\hspace{3.8cm}  \= 3-point\hspace{0.8cm}     \= 3-point+Gaussian \\
$S(I)$         \> 7.9\,Jy \> 4.6\,Jy \\
$S(\rm{SMA1})$ \> 7.1\,Jy \> 3.5\,Jy \\
$S(\rm{Hot Core (HC)})$  \> 3.1\,Jy \> 2.4\,Jy \\
Gaussian       \>        \> 178\,Jy (size $6.8''\times 2.7''$)
\end{tabbing}

Including the 25\% calibration uncertainty, the 440\,$\mu$m
fluxes measurements with conservative estimates of the error budget
are:
\begin{eqnarray*}
3.5<S(I)<9.9 & \Rightarrow & S(I)\sim 6.7\pm 3.2\,\rm{Jy} \\
2.6<S(\rm{SMA1})<8.9 & \Rightarrow & S(\rm{SMA1})\sim 5.75\pm 3.15\,\rm{Jy}\\
1.8<S(\rm{HC})<3.9 & \Rightarrow & S(\rm{HC})\sim 2.85\pm 1.05\,\rm{Jy}
\end{eqnarray*}

The errors are not to be understood as 1$\sigma$ or 3$\sigma$ values,
but they rather give the extreme values of what the fluxes of the
three sources could be. As outlined above, the line contamination of
our continuum dataset is very low.

\subsubsection{The large-scale structure}

Obviously, from our data we cannot constrain well the large-scale flux
distribution. However, the fitted 178\,Jy are a lower limit to the
large-scale flux since we have no data below 28\,k$\lambda$, and
\citet{johnstone1999} measured a 450\,$\mu$m single-dish peak flux of
490\,Jy with SCUBA at $7.5''$. Furthermore, we investigated the
influence of a large source with a flux of a few hundred Jy on our
observations. Modeling different source sizes and {\it observing} them
with the given uv-coverage, the size-scale of the large-scale emission
has to exceed $11''$. Larger sources are filtered out by the
observations and do not distort the observed continuum image
significantly, whereas smaller sources with large flux would
completely dominate the observations. The scale of $11''$ corresponds
approximately to the size-scales theoretically traceable by the
shortest baselines of 28\,k$\lambda$ (corresponding to approximately
$9''$ scales).

\subsubsection{Source I} 
\label{source_i}

Prior to these observations, the SED of source {\it I} could be fitted
well by two very different models, either as the result of optically
thick proton-electron free-free emission up to 100\,GHz plus dust
emission that accounts for most of the submm flux, or H$^-$ free-free
emission that gives rise to a power-law spectrum with a power-law
index of $\sim 1.6$ \citep{beuther2004g}. Since the two models vary
strongly at higher frequencies, an accurate 440\,$\mu$m continuum flux
measurement was of utmost importance in discriminating between these
models.

\begin{figure}
\begin{center}
\includegraphics[angle=-90,width=9cm]{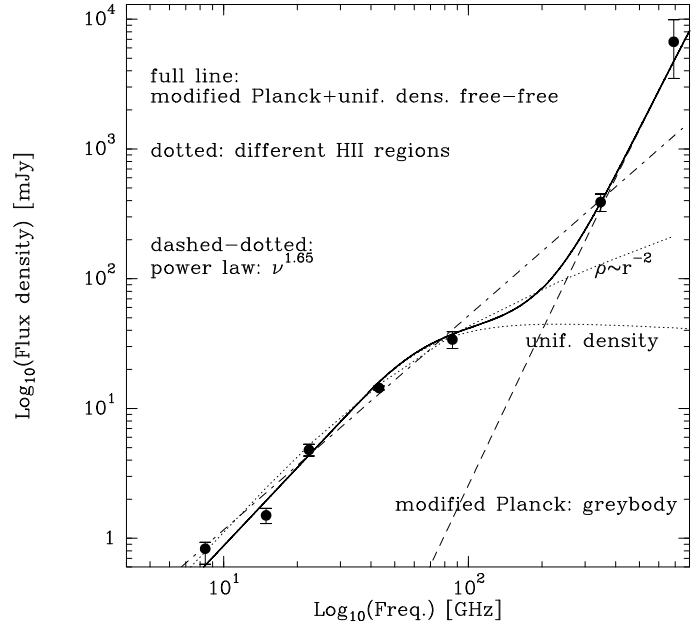}
\end{center}
\caption{The SED of source {\it I}. The measured fluxes are labeled as
filled circles with error bars, and the various lines show different
attempts to fit the data as labeled within the plot. The 15, 22 and
86\,GHz fluxes are taken from \citet{plambeck1995} and references
therein, the 8 and 43\,GHz measurements are more recent values
(consistent with the error bars of the previous measurements by
\citet{menten1995}, M.J. Reid et al.~in prep.), the 345\,GHz data
point is from \citet{beuther2004g}, and the 690\,GHz point is from
this work.}
\label{sed}
\end{figure}

Figure \ref{sed} shows the complete SED of source {\it I} from 8 to
690\,GHz. The measured 690\,GHz flux fits the model of a
proton-electron free-free plus dust emission spectrum well, but not
the model of the power-law spectrum expected for H$^-$ free-free
emission. Therefore, the 440\,$\mu$m continuum data can discriminate
between the two models, and the SED of source {\it I} turns out to be
a rather typical SED for deeply embedded protostellar sources.  The
important result here is that the turn-over frequency (i.e., the
transition from optically thick to optically thin emission) is much
higher in source {\it I} than in other protostars where it is
typically in the range of 10 to 40\,GHz.

\subsubsection{SMA1} 

The observations show that SMA1 is not just an extension of the hot
core, but is an independent protostellar source. Based on PdBI
HC$_3$N(10--9) observations of vibrational excited lines,
\citet{devicente2002} predicted an additional massive protostellar
source in the region approximately $1.8''$ south of source {\it I},
which is the location of SMA1. Only two flux measurements exist for
SMA1: at 865\,$\mu$m ($S\sim 0.36\pm 0.07$\,Jy,
\citealt{beuther2004g}) and at 440\,$\mu$m ($S\sim 5.75\pm 3.15$\,Jy).
The spectral index between these two bands is $S \propto
\nu^{\alpha}$ with $\alpha \sim 4.0^{+0.9}_{-1.4}$. In spite of the
broad range of potential $\alpha$, the data are consistent with a
typical dust opacity index $\beta = \alpha - 2$ in star-forming
regions of 2. The non-detection of SMA1 at cm and infrared wavelengths
suggests that it may be one of the youngest sources of the evolving
cluster.

\subsubsection{The Hot Core}
\label{hotcore}

The hot core flux values we derived above are probably more uncertain
than those from source {\it I} and SMA1 because the hot core is only
weakly discernible from the rest of the image in the original data as
well as the model. Furthermore, the response of the continuum and line
emission to the poor uv-sampling is not the same in interferometric
studies, and we cannot entirely exclude that the weakness of the hot
core in the 440\,$\mu$m data is caused by this poor uv-sampling and
the resulting imaging problems. Keeping in mind these uncertainties,
we nevertheless can use the existing data to estimate a spectral index
between the previous 865\,$\mu$m measurement and this new data. The
flux measurements in the two bands toward the hot core peak position
($S(865\mu\rm{m})\sim 0.54\pm 0.11$\,Jy, \citealt{beuther2004g}, and
$S(440\mu\rm{m})\sim 2.85\pm 1.05$\,Jy) result in a range of potential
spectral indices $S \propto \nu^{\alpha}$ with $\alpha \sim
2.4^{+0.8}_{-0.9}$. These values are considerably lower than for
source {\it I} and SMA1.

Single-dish spectral index studies toward the Orion A molecular cloud
qualitatively find a somewhat lower spectral index toward Orion-KL
than toward most of the rest of the Orion A molecular cloud
\citep{lis1998,johnstone1999}, but the two studies do not agree
quantitatively. \citet{lis1998} find a dust opacity index $\beta \sim
1.8$ corresponding in the Rayleigh-Jeans limit to a spectral index
$\alpha \sim 3.8$, whereas \citet{johnstone1999} find a spectral index
$\alpha \sim 2.2$.  Nevertheless, considering the large number of
potential uncertainties (e.g., calibration, different beam sizes,
line contamination, etc.), the qualitatively lower spectral index
$\alpha$ toward Orion-KL compared with its surroundings appears a
reliable result from the single-dish studies.  

In contrast to source {\it I} and SMA1, where the spectral indices are
both consistent with typical values of $\alpha$ around 4 in
star-forming regions, the lower spectral index toward the hot core
is intriguing. Furthermore, lower-resolution observations show
that on larger spatial scales the data are dominated by the apparently
more extended hot core emission (e.g.,
\citealt{plambeck1995,blake1996,beuther2004g}). Therefore, the
different spectral indices indicate that the lower single-dish
spectral index may be due to hot core emission distributed over larger
spacial scales, whereas the protostellar sources {\it I} and SMA1
exhibit spectral indices more typical of those from other star-forming
regions.

\subsection{Spectral line emission}
\label{line}

We detected 24 spectral lines over the entire bandpass of 4\,GHz
(Fig.~\ref{spectra} \& Table \ref{lines}). This is a significantly
lower line number than the $\sim 150$ lines over the same spectral
width at 348\,GHz. The lower sensitivity of the 690\,GHz data compared
with the 345\,GHz observations accounts for this difference to some
degree. Furthermore, at 690\,GHz each line of a given $\Delta v$
covers twice the frequency range than at 345\,GHz, and thus strong
lines dominate a broader part of the spectrum than at lower
frequencies. Furthermore, to first order the observing frequency is
proportional to the average energy levels of observable lines in that
frequency range, and the lower number of line detections at 690\,GHz
indicates that the bulk of the gas is at lower temperatures (see
temperature discussion in \S\ref{ch3cn}). Nevertheless, the energy
levels of the 690\,GHz lines (Table \ref{lines}) show that a very warm
gas component has to be present in the region as well. Temperature
estimates based on the CH$_3$CN$(37_K-36_K)$ are derived below (\S
\ref{ch3cn}). Comparing these spectra with the single-dish spectra
toward the same region \citep{schilke2001,harris1995}, we find a large
number of lines previously not detected in this region. Line
identifications were done first using the line survey by
\citet{schilke2001}, and then cross-checking the identifications with
the molecular databases at JPL and CDMS
\citep{poynter1985,mueller2001}. It has to be noted that the
unambiguous identification of spectral lines becomes more difficult at
higher frequencies, which is discussed in \S \ref{ul}.

Table \ref{lines} also shows the line-widths derived from Gaussian
fits to the vector-averaged spectra on a short baseline of 25\,m
(Fig.~\ref{spectra}). Since the measured line-widths are associated
with gas from the whole emission in the primary beam, they do not
represent the line-width toward a selected position but rather an
average over the whole region. The derived values for $\Delta v$ range
between 4.8 and 16.4\,km\,s$^{-1}$. We could not fit the CO(6--5) line
due to the missing flux. While some of the broader line-widths may
be due to the outflows in the region, this is probably not the main
reason for most of the lines since the spectra often resemble a
Gaussian profile. It seems more likely that different velocity
components as well as large intrinsic line-widths toward each position
due to turbulence and/or internal motion cause a reasonable fraction
of the line-width. For a discussion of the CH$_3$CN line-widths, see
\S\ref{ch3cn}.

Figure \ref{maps} presents the images of the velocity-integrated
emission of a few of the identified molecular lines in the
bandpass. As discussed in \S\ref{obs} and \S\ref{cont}, imaging of the
emission turns out to be difficult, as evidenced by the strong
negative side-lobes in Fig.~\ref{maps}. Since our primary beam is
rather small ($18''$) and the imaging gets worse towards the edge of the
primary beam, we consider only emission at the centers of the maps to
be reliable. All line images in Fig.~\ref{maps} show considerable
emission toward source {\it I}, SMA1 and the hot core peak
position. C$^{33}$S(14--13) peaks toward SMA1 with lower level
emission from the hot core and source {\it I}. In contrast to this,
the line images from CH$_3$CN$(37_2-36_2)$, CH$_3$OH$(22_1-21_2)$ and
SO$_2(35_{3,33}-34_{2,32})$ all show emission peaks from source {\it
I}, SMA1 and the hot core peak position. The
HNCO$(31_{2,29}-30_{2,28})$ image exhibits emission peaks associated
with source {\it I} and SMA1, and only lower-level emission from the
hot core position. A striking difference between this 440\,$\mu$m line
images and our previous study in the 865\,$\mu$m band
\citep{beuther2005a} is that the higher excited lines in the
440\,$\mu$m band show line emission from source {\it I}. In the
865\,$\mu$m band, we found mainly SiO emission from source {\it I},
other species appear to avoid the source in the lower frequency
band. This indicates the presence of high temperatures in the close
environment of source {\it I}.

We could not produce reliable emission maps for the CO(6--5) and
H$^{13}$CN(8--7) lines. While this is less of a surprise for CO, which
is known to be extended and thus difficult to image
interferometrically, H$^{13}$CN was expected to be far more compact,
comparable to many of the other imaged lines. However, comparing all
lines in our frequency setup (excluding only the CO(6--5) line), the
H$^{13}$CN(8--7) line has one of the broadest linewidths ($\Delta v
\sim 16.2$\,km\,s$^{-1}$), and thus may be more strongly affected by
the molecular outflows present than initially expected. This could
explain our inability to properly image the H$^{13}$CN(8--7) emission.

\subsection{CH$_3$CN$(37_K-36_K)$: temperature estimates and line-widths}
\label{ch3cn}

Since we observed the $K=0,1,2,3$ lines of the CH$_3$CN$(37_K-36_K)$
lines (the $K=0,1$ lines are blended), we can use this $K$-series
for temperature estimates of the warm gas component in the
regions. Figures \ref{maps} \& \ref{ch3cn_spectra} show the integrated
CH$_3$CN$(37_2-36_2)$ emission map and spectra of all $K$-components
toward the CH$_3$CN peak positions associated with source {\it I},
SMA1, and the hot core. 

Using the XCLASS superset to the CLASS software (Peter Schilke,
priv.~comm.), we produced model spectra in local thermodynamic
equilibrium (Fig.~\ref{ch3cn_spectra}). This software package uses the
line catalogs from JPL and CDMS
\citep{poynter1985,mueller2001}. Because the $K=3$ component has
double the statistical weight than the $K=0,1,2$ lines, one expects in
the optical thin case larger line intensities for the $K=3$
transition. The optically thin model spectra reproduce the relative
line intensities well toward SMA1 and the hot core. Toward source {\it
I}, the modeled $K=3$ intensity is a bit larger than observed
indicating increasing optical depth there. The upper energy levels
$E_u$ of the various K-ladder lines range only between 621 and 685\,K
(Table \ref{lines}), and the data are not very sensitive to the
temperature. We find reasonable model spectra in the 600\,K regime
with a large potential range of temperatures of $\pm 200$\,K. The
CH$_3$CN model column densities we used are of the order a few times
$10^{15}$\,cm$^{-2}$, and the line-width adopted for the modeling was
6.5\,km\,s$^{-1}$. In spite of the large errorbars, the temperatures
traced by these high-energy lines are, as expected, higher than the
highest temperatures of 350\,K previously derived from our CH$_3$OH
multi-line analysis in the 865\,$\mu$m band \citep{beuther2005a}. We
attribute this temperature difference mainly to the various gas
components at different temperatures and densities, and to the
different excitation energies of the spectral lines in the 690 and
337\,GHz bands.

Furthermore, we can compare the modeled CH$_3$CN$(37_K-36_K)$
line-widths of the order 6.5\,km\,s$^{-1}$ with the previous SMA
observations of the CH$_3$CN$(19_8-18_8)$ line at 348\,GHz
\citep{beuther2005a}. The measured CH$_3$CN$(19_8-18_8)$ from SMA1
and the hot core are 4.6 and 4.1\,km\,s$^{-1}$,
respectively. \citet{beuther2005a} found only very weak
CH$_3$CN$(19_8-18_8)$ toward source {\it I}. Although the excitation
temperatures $E_u$ of the CH$_3$CN$(37_2-36_2)$ and the
CH$_3$CN$(19_8-18_8)$ lines are approximately the same (649 and
624\,K, respectively), it is interesting that the line-width toward
the CH$_3$CN$(37_2-36_2)$ is larger. Since the critical densities
between the CH$_3$CN$(37_K-36_K)$ and CH$_3$CN$(19_K-18_K)$ series
vary by about one order of magnitude ($\sim 10^8$ and $\sim
10^7$\,cm$^{-3}$, respectively), this indicates that the higher
density gas CH$_3$CN$(37_K-36_K)$ lines are likely subject to
significantly more turbulent motions. As already pointed out by
\citet{sutton1986}, radiative excitation is unlikely to significantly
affect the populations of the vibrational ground state transitions,
even for high J-levels.

\subsection{Unidentified lines (UL)}
\label{ul}

Fourteen out of the 24 detected spectral lines remain unidentified or
only tentatively identified. Such a high percentage of unidentified
lines ($\sim 58\%$) has rarely been reported, to our knowledge only
recently toward SgrB2 \citep{friedel2004}. The previous single-dish
line surveys of Orion-KL report percentages of unidentified lines
between 8 and 15\%
\citep{schilke1997b,schilke2001,comito2005}. Orion-KL is known to have
a number of different velocity components approximately between 2 and
9\,km\,s$^{-1}$ (e.g., \citealt{genzel1989}), and we cannot determine
at which velocity an unidentified line is beforehand. Hence, at
690\,GHz the frequency range of potential molecular lines between 2
and 9\,km\,s$^{-1}$ is 16\,MHz, compared to 8\,MHz at 345\,GHz
($\Delta\nu = \Delta v \times
\nu /c$). Furthermore, going to higher frequencies, astronomical
spectral line studies are rare and identifications that rely only on
laboratory work are difficult. Hence, identifying spectral lines gets
more complicated at higher frequencies.

Figure \ref{unidentified} presents some of the integrated line
emission maps of the ULs. Since these lines are on average of lower
intensity, the signal-to-noise ratio in the maps is worse than for the
identified lines in Fig.~\ref{maps}. The rest-frequencies listed for
the ULs in Table \ref{lines} (and shown in Fig.~\ref{unidentified})
correspond to the lines being set to a $v_{\rm{lsr}}$ of
5\,km\,s$^{-1}$, implying the potential range of frequencies for each
line as discussed above. In principle, it should sometimes be possible
to associate unidentified lines with molecular families (e.g.,
oxygen-, nitrogen-, or sulphur-bearing species) based on the spatial
distribution of the gas (see, e.g., the previous 865\,$\mu$m Orion
line studies by \citealt{beuther2005a}). However, since the quality of
the images is rather poor, this is a difficult task for this
dataset. Nevertheless, we tried to associate some unidentified lines
with potential molecular transitions. Table \ref{lines} lists a few
tentative candidate lines for the ULs, which, however, are only
suggestions and not real identifications.

\section{Discussion and Conclusions}

The presented line and continuum data of Orion-KL in the
440$\mu$m/690GHz band show the power of high-spatial-resolution
studies in the submm wavelength bands. The measured continuum flux
from source {\it I} allows us to differentiate between various
previously proposed physical models: Source {\it I} appears to be a
rather ``normal'' protostellar object with a SED fitted by a
two-component model of proton-electron free-free emission below
100\,GHz plus a dust component contributing to the flux in the submm
bands. Furthermore, the source SMA1 becomes more prominent at higher
frequencies and is clearly distinguishable from the hot core
emission. Since SMA1 is detected neither at cm nor at infrared
wavelengths, it may be one of the youngest sources of the evolving
cluster. The only weak detection of the hot core at 440\,$\mu$m is
puzzling. Although it might be a real feature of the new
high-spatial-resolution 440\,$\mu$m continuum data, we cannot entirely
exclude that it is caused by the poor uv-sampling and resulting
imaging problems. Keeping the uncertainties in mind, we find a lower
spectral index toward the hot core compared with source {\it I} and
SMA1. This is consistent with lower spectral indices found toward
Orion-KL in lower-spatial-resolution single-dish observations that are
dominated by the hot core emission.

The spectral line maps trace a warm gas component at the center of the
region, mainly confined to the sources {\it I}, SMA1 and the hot core
peak position.  Temperature estimates based on the
CH$_3$CN$(37_K-36_K)$ K-ladder ($K=0...3$) lines indicate a warm gas
component in the regime of $600\pm 200$\,K. The number of unidentified
lines in the given setup is large. Potential reasons are discussed, of
which the two main ones are likely the large potential spread in
velocity, and thus frequency, of the ULs, and the less explored submm
wavelength band.

\acknowledgments{We would like to thank Peter Schilke for providing the
XCLASS software to model the CH$_3$CN spectra. Thanks very much also to
Darek Lis for discussing spectral indices in Orion-KL. We also
appreciate the helpful comments from the referee. H.B. acknowledges
financial support by the Emmy-Noether-Program of the Deutsche
Forschungsgemeinschaft (DFG, grant BE2578/1).}


\clearpage

\begin{figure}
\begin{center}
\end{center}
\caption{Submillimeter continuum images of Orion-KL. Positive emission is
always shown in gray scale with contours, negative features~-- due to
missing short spacings and the calibration/imaging difficulties
discussed in \S\ref{obs}~-- are presented in dashed contours. The {\bf
left} panel shows the previously published 348\,GHz continuum data
\citep{beuther2004g}. The contours start at the 2$\sigma$ level of
70\,mJy\,beam$^{-1}$ and continue in 2$\sigma$ steps. The main sources
in the region are labeled by name. The {\bf middle} panel shows an
image of the same region, with a limited range of uv-data between $30$
and $160$\,k$\lambda$. The contouring is the same as in the left
panel, and the stars mark the four sources identified in the left
panel. The {\bf right} panel shows the new 690\,GHz continuum data
covering a uv-range between 30 and 160\,k$\lambda$ again (blanking the
only bin below 30\,k$\lambda$ at 28\,k$\lambda$). The contouring is
done in 10 to 90\% steps from the peak emission of 10.3\,Jy, and the
stars again mark the sources identified in the left panel. The
synthesized beams are presented at the bottom-left of the panels. The
different synthesized beams in the middle and right panel result from
a different uv-sampling, although we applied the same uv-limits.}
\label{continuum}
\end{figure}

\begin{figure}
\begin{center}
\end{center}
\caption{Original image and model-fits to the 690\,GHz continuum data.
{\bf 1.)} Original image, {\bf 2.)} 2 point source model, {\bf 3.)} 2
point source + Gaussian model, and {\bf 4.)} 3 point source + Gaussian
model. The contouring is done in $\pm 10$ to $\pm 90$\% steps from the
peak emission of the three images. The peak values are 10.3, 7.7, 10.2
and 10.6\,Jy for panels 1 to 4, respectively. The stars again mark the
sources identified in the left panel of Figure \ref{continuum}. The
synthesized beams are presented at the bottom-left of each panel.}
\label{model}
\end{figure}

\begin{figure}
\begin{center}
\end{center}
\caption{Integrated line emission maps after continuum subtraction: The 
top-left panel shows for comparison again the continuum image as
presented in Fig.~\ref{continuum}, and the other 5 panels show
representative molecular line maps as labeled within each
panel. Positive emission is presented in gray scale and full contours,
and negative emission~-- due to missing short spacings~-- is shown in
dashed contours. The contours always range from 10 to 90\% of the peak
emission in each image. The synthesized beams are shown at the
bottom-left of each panel.}
\label{maps}
\end{figure}

\begin{figure}
\begin{center}
\includegraphics[angle=-90,width=9cm]{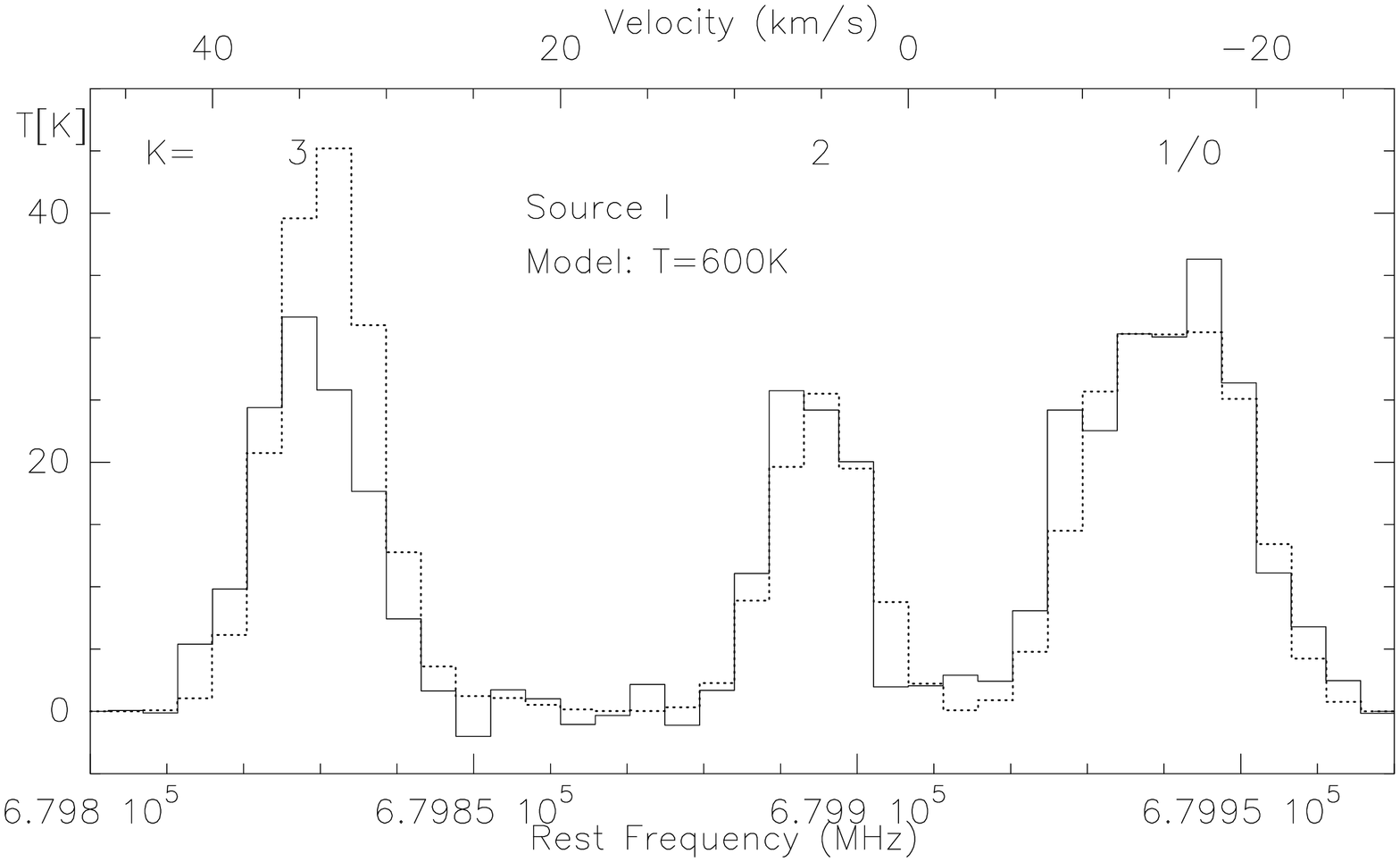}\\
\includegraphics[angle=-90,width=9cm]{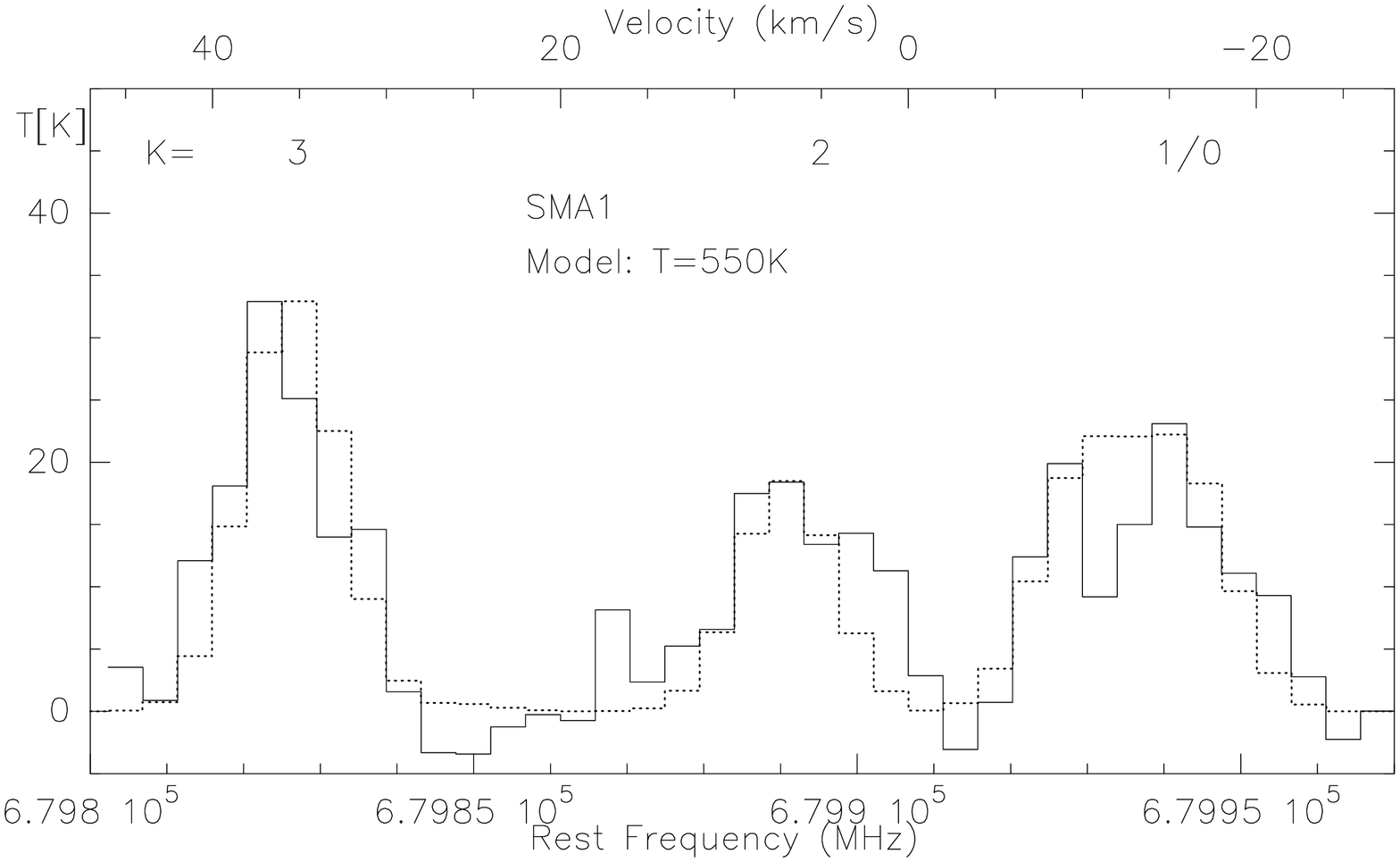}\\
\includegraphics[angle=-90,width=9cm]{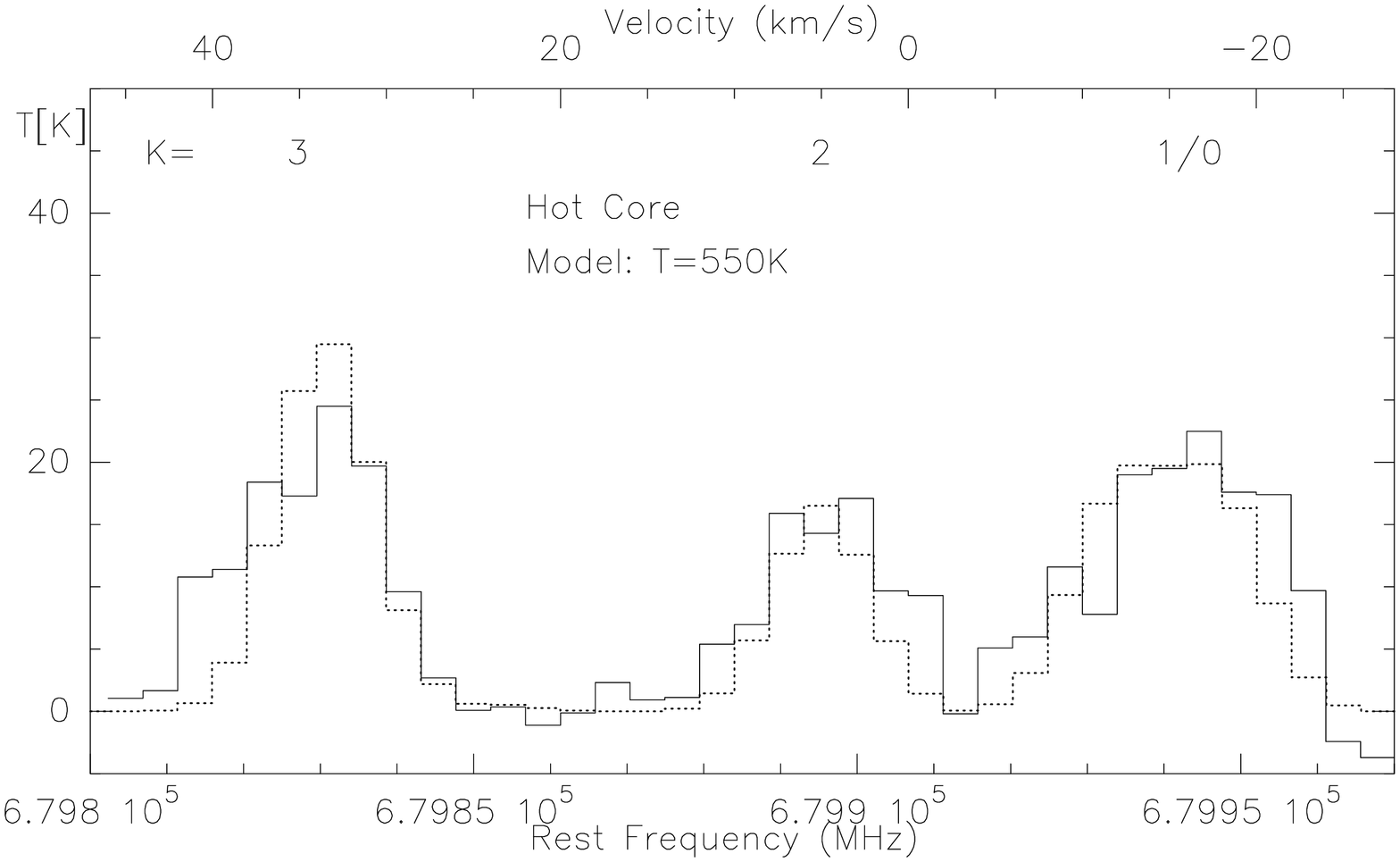}
\end{center}
\caption{CH$_3$CN$(37_K-36_K)$ spectra taken toward the CH$_3$CN peak
positions associated with source {\it I}, SMA1 and the hot core (see
Fig.~\ref{maps}). The full lines show the data, and the dotted lines
LTE model spectra (adopted temperatures are given in each panel). The
data are not very sensitive to the temperatures and can be modeled
with $T$ varying approximately $\pm 200$\,K.}
\label{ch3cn_spectra}
\end{figure}

\begin{figure}
\begin{center}
\end{center}
\caption{Integrated line emission maps of some unidentified lines 
after continuum subtraction: positive emission is presented in gray
scale and full contours, and negative emission~-- due to missing short
spacings~-- is shown in dashed contours. The contours always range
from 10 to 90\% of the peak emission in each image. The rest-frequency
for each line set to a $v_{\rm{lsr}}=5$\,km\,s$^{-1}$ is shown at the
top-left of each map. The symbols are the same as in the previous
Figures, and the synthesized beams are shown at the bottom-left of
each panel.}
\label{unidentified}
\end{figure}

\clearpage

\begin{deluxetable}{lrrrr}
\tablecaption{Observed lines \label{lines}}
\tablewidth{0pt}
\tablehead{
\colhead{$\nu^a$} & \colhead{line} & \colhead{$E_u$} & \colhead{$\Delta v^b$} & \colhead{Tentative?$^c$} \\
\colhead{[GHz]} & \colhead{} & \colhead{[K]} & \colhead{[km\,s$^{-1}$]} & \colhead{} \\
}
\startdata
679.781 & C$^{33}$S(14--13) & 247 & 7.1 & \\
679.831 & CH$_3$CN$(37_3-36_3)$ & 688 & 8.3 & \\
679.895 & CH$_3$CN$(37_2-36_2)$ & 649 & 8.3 & \\
679.934 & CH$_3$CN$(37_1-36_1)$ & 628 & 13.4$^f$ & \\
679.947 & CH$_3$CN$(37_0-36_0)$ & 621 & 13.4$^f$ & \\
680.210 & UL$^d$ & & 7.1 & C$_2$H$_3$CN$(72_{4,69}-71_{4,68})$@680.211 \\
        &        & &      & OCS$(56-55)$@680.213 \\
680.485 & UL & & 13.8$^f$ & C$_2$H$_5$CN$(78_{4,75}-77_{4,74})$@680.481\\
680.500 & UL & & 13.8$^f$ & C$_2$H$_5$CN$(78_{3,75}-77_{3,74})$@680.507\\
680.804 & CH$_3$OH$(22_1-21_2)A^-$ & 606 & 12.6 & \\
680.890 & HNCO$(31_{2,29}-30_{2,28})$ & 696 & 16.4 & \\ 
681.086 & UL & & 4.8 & CH$_3$CN$(37_7-36_7)v_8=1$@681.087\\
        &    & &      & CH$_3$OCH$_3(19_{6,13}-18_{5,14})$@681.090\\
681.150 & UL & & 5.5 & C$_2$H$_5$OH$(56_{21,35}-56_{21,35})$@681.152\\
681.218 & UL & & 5.2 & CH$_3$CN$(37_9-36_9)v_8=2$@681.216\\
        &    & &      & C$_2$H$_5$CN$(70_{8,63}-71_{3,68})$@681.224 \\   
681.263 & UL & & 10.4 & HCOOCH$_3(35_{8,28}-35_{4,31})$@681.265\\
        &    & &      & SO$_2$$(68_{9,59}-68_{8,60})$@681.266\\
        &    & &      & $^{33}$SO$(16-15)$@681.269\\
681.328 & UL & & 6.3 & CH$_3$CN$(37_1-36_1)v_8=2$@681.329\\
681.440 & UL & & 8.6 & CH$_3$CN$(37_5-36_5)v_8=1$@681.438\\
681.589 & UL & & 8.7 & CH$_3$CN$(37_7-36_7)v_8=2$@681.590\\
681.674 & SO$_2(35_{3,33}-34_{2,32})$ & 598 & 11.4 & \\
681.733$^e$ & UL & & & \\
689.840 & UL & & 5.2 & D$_2$CO$(20_{2,18}-20_{2,19})$@689.840\\
        &    & &      & C$_2$H$_3$CN$(21_{6,15}-22_{4,18})$@689.842\\
        &    & &      & SiN$(16-15)$@689.842\\
        &    & &      & C$_2$H$_5$CN$(28_{10,19}-27_{9,18})$@689.842\\
690.349 & UL & & 7.0 & \\
690.551 & H$^{13}$CN(8--7) & 149 & 16.2 & \\
690.890 & UL & & 6.0 & HCOOCH$_3$$(27_{7,21}-26_{6,20})$@690.891\\
        &    & &      & C$_2$H$_3$CN$(72_{30,42}-71_{30,41})$@690.892\\
        &    & &      & C$_2$H$_5$CN$(23_{11,12}-22_{10,13})$@690.892 \\
691.473 & CO(6--5) & 116 & ?$^g$ &
\enddata
\tablenotetext{a}{\footnotesize The frequency uncertainties are below
1\,MHz, mostly even below 0.1\,MHz. The frequencies for the ULs are
derived via setting the $v_{\rm{lsr}}$ to 5\,km\,s$^{-1}$.}
\tablenotetext{b}{\footnotesize FWHM of Gaussian fits to the
vector-averaged spectra of the short 25\,m baseline (see
Fig.~\ref{spectra}).} \tablenotetext{c}{\footnotesize Tentative
molecular transition identifications which are still very uncertain!}
\tablenotetext{d}{\footnotesize UL: Unidentified line.}
\tablenotetext{e}{\footnotesize This line is at the edge of the
bandpass and thus only partly detected. Therefore, the frequency is
even more uncertain.}  \tablenotetext{f}{\footnotesize Line blend.}
\tablenotetext{g}{\footnotesize Due to missing flux, a line fit does
not make sense.}
\end{deluxetable}

\end{document}